\documentclass{ar-1col}

\usepackage[numbers]{natbib}
\usepackage{amsmath}
\usepackage{amsfonts}
\usepackage{bm}
\usepackage{url}
\newcommand{\Vk}{\vec{k}}
\newcommand{\vr}{\boldsymbol{r}}
\newcommand{\vq}{\boldsymbol{q}}
\newcommand{\VA}{\vec{A}}
\newcommand{\VB}{\vec{B}}
\newcommand{\Vd}{\vec{d}}

\jname{Xxxx. Xxx. Xxx. Xxx.}
\jvol{AA}
\jyear{YYYY}
\doi{10.1146/((please add article doi))}

\begin{document}

\markboth{Fischer et al.}{Locally Noncentrosymmetric Superconductors}

\title{Superconductivity and Local Inversion-Symmetry Breaking}

\author{Mark H. Fischer$^1$ Manfred Sigrist,$^2$, Daniel F. Agterberg,$^3$ and Youichi Yanase$^4$
\affil{$^1$Department of Physics, University of Zurich, 8057 Zurich, Switzerland; email: mark.fischer@uzh.ch}
\affil{$^2$Institute for Theoretical Physics, ETH Zurich, 8093 Zurich, Switzerland; email: sigrist@itp.phys.ethz.ch}
\affil{$^3$Department of Physics, University of Wisconsin-Milwaukee, Milwaukee, Wisconsin 53201, USA; email: agterber@uwm.edu}
\affil{$^4$Department of Physics, Graduate School of Science, Kyoto University, Kyoto 606-8502, Japan; email: yanase@scphys.kyoto-u.ac.jp}}

\begin{abstract} 
Inversion and time reversal are essential symmetries for the structure of Cooper pairs in superconductors. The loss of one or both leads to modifications to this structure and can change the properties of the superconducting phases in profound ways. Lacking inversion, superconductivity in noncentrosymmetric materials has become an important topic, in particular, in the context of topological superconductivity as well as unusual magnetic and magneto-electric properties.
Recently, crystal structures with local, but not global inversion-symmetry breaking have attracted attention, as superconductivity can exhibit phenomena not naively expected in centrosymmetric materials. After introducing the concept of locally noncentrosymmetric crystals and different material realizations, we discuss consequences of such local symmetry breaking on the classification, the expected and, in parts, already observed phenomenology of unconventional superconductivity, and possible topological superconducting phases.
\end{abstract}

\maketitle

\section{INTRODUCTION}
The importance of time-reversal and inversion symmetry for superconductivity was realized early on: These two symmetries guarantee the degeneracy of electrons at the Fermi level with opposite momentum and thus, a weak-coupling instability towards superconductivity~\cite{anderson:1959, anderson:1984}.
For (conventional) spin-singlet superconductivity, time-reversal symmetry (TRS) is sufficient for this so-called Cooper instability, while additional inversion symmetry allows for general spin-triplet superconductivity.
A further consequence of inversion symmetry is that one can classify order parameters into even and odd parity. For single-band systems, this prohibits mixing of spin-singlet and spin-triplet pairing channels.

While time-reversal symmetry and inversion allow for superconducting instabilities, breaking or the absence of these symmetries has direct consequences for the superconducting state. Indeed, removing time-reversal symmetry by an external magnetic field, magnetic impurities, or (ferro-) magnetic order substantially weakens or even suppresses superconductivity in the spin-singlet channel. 
Together with the clear distinction of even and odd superconducting order parameters in systems with inversion, the response to a magnetic field thus allows to distinguish spin-singlet and spin-triplet superconductors, an important interplay of the two symmetries.

Breaking of inversion, or rather superconductivity in a crystal without a center of inversion, has attracted much interest since the discovery of superconductivity in the noncentrosymmetric CePt$_{3}$Si in 2004~\cite{bauer:2004} and plays an important role in this review. Among the most intriguing features of such noncentrosymmetric superconductors are:
\begin{enumerate}

\item The unusual response of even conventional superconductors to magnetic fields \cite{gorkov:2001,frigeri:2004,frigeri:2004-2}, including substantially enhanced Pauli limiting fields and a non-vanishing spin-susceptibility down to zero temperature.
\item Parity mixing of Cooper-pair states~\cite{edelshtein:1989, gorkov:2001,frigeri:2004} often called singlet-triplet mixing.
\item A finite-momentum pairing state, often called the helical superconducting phase, induced by a magnetic field~\cite{mineev:1994, kaur:2005a}.
\item The appearance of topological superconducting phases~\cite{sato:2009b, sato:2009c,Schnyder:2015, RevModPhys.88.035005}. 
\end{enumerate}
Initial investigations into local inversion-symmetry breaking were motivated by asking whether the above features associated with noncentrosymmetric superconductors could be observed when inversion symmetry is globally restored but locally broken.

The importance of local symmetry breaking has a long history in condensed matter physics. In particular, novel phenomena can emerge for the situation, where symmetries are broken locally in a sublattice, such that the symmetry breaking can be undone by interchange of the sublattices.
Indeed, Baltensperger and Straessler demonstrated in 1962 that spin-singlet superconductivity and antiferromagnetism can coexist for an appropriate pair structure~\cite{baltensperger:1962}. Specifically, staggered moments break time-reversal symmetry only on sublattices resulting in a spin-singlet pair wave function with significant amplitudes only when the two electrons are on different sublattices.

Similarly, a crystal structure can locally have a reduced symmetry. While taking the local crystal symmetry into account for the description of localized magnetic moments goes back to Dzyaloshinsky and Moriya~\cite{dzyaloshinsky:1958, moriya:1960} and allows for an explanation of the phenomena of weak ferromagnetism in antiferromagnetic systems (canted AFM), only recently have people taken similar spin-orbit-coupling effects into account when describing itinerant electron systems~\cite{fischer:2010, fischer:2011b, zhang:2014}. These effects have, further, attracted a lot of attention in the magnetism community in the context of antiferromagnetic spintronics~\cite{zelezny:2014, wadley:2016}, piezoelectronics~\cite{watanabe2017,shiomi2019}, and nonlinear optics~\cite{Zhang2019,Ahn2020,watanabe2021}.

In this review, we discuss consequences of the local breaking of inversion symmetry on superconductivity. After a general discussion of the symmetry of crystal lattices, we discuss several classes exemplifying how inversion can be broken locally. We then review the consequences of local inversion-symmetry breaking on the normal-state microscopic Hamiltonian, the classification of the superconducting order parameter, and the physical consequences on the superconducting state---emphasizing similarities and differences with respect to globally inversion-symmetry-broken (noncentrosymmetric) superconductors. 
Finally, we conclude with a discussion of peculiarities of these systems going beyond the phenomena known from noncentrosymmetric superconductors and routes to enhance the role of local inversion-symmetry breaking.

\section{BASIC THEORETICAL CONCEPTS}\label{sec:basics}

\subsection{Lattice Symmetries and Group Theory} 
For a crystal lattice, the spatial symmetry transformations, namely point group and translation symmetries, that map all atomic sites of the crystal onto themselves form a group, the space group. For non-magnetic crystals, there are 230 distinct space groups. The pure translations form the translation group and encode the periodicity of the lattice with respect to a unit cell. Point group operations include rotations, mirror operations, and combinations thereof, such as inversion~\cite{dresselhaus:2007}. The space group is crucial for the description of an electronic system both in the normal and the superconducting state: For the normal state, the translation group leads to the description of electronic states by means of the Bloch wave functions with quasi-momentum in the Brillouin zone (BZ) and degeneracies within the BZ governed by the point group. For the superconducting state, we classify the superconducting order parameters according to their irreducible representation of the point group, a crucial ingredient for understanding the physics of unconventional superconductivity~\cite{sigrist:1991}.

While the crystal is mapped onto itself under all elements of the point group, the environment of a generic atomic site in the lattice is in general less symmetric. All symmetry-distinct sites in the unit cell, captured by the concept of Wyckoff positions, have their own symmetry group, the so-called site symmetry group. The site symmetry group is a subgroup of the point group with transformations that leave the specific site invariant. Consequently, we can construct a sublattice from such sites with lower symmetry, such that this sublattice lacks symmetries of the full crystal structure, such as inversion. 

When a crystal structure comprises two sublattices that lack inversion symmetry but the point group of the crystal contains inversion symmetry (which exchanges the sublattices), we call such crystal structures \emph{locally noncentrosymmetric}. If for some reason the sublattices related by inversion symmetry are only weakly coupled, in other words there is a natural separation of the sublattices, it can be instructive to discuss the individual sublattices and their symmetry first, before combining the sublattices into the full crystal structure. Consequently, the resultant physics will be reminiscent of two copies of a noncentrosymmetric material---in the spintronics community this led to the notion of `hidden spin-polarization'~\cite{zhang:2014}. It is typically in this situation when new physics emerges, leading to the clearest manifestations of locally noncentrosymmetric superconductivity. 

\begin{figure}[tt]
\includegraphics{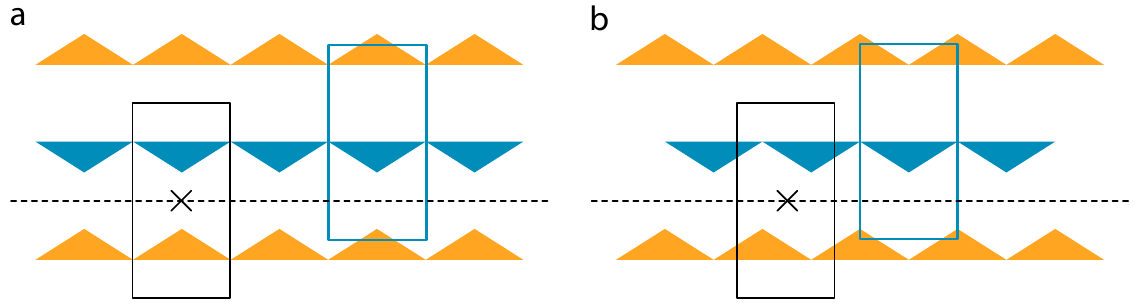}
\caption{Locally noncentrosymmetric structures with the black box denoting an inversion-symmetric unit cell. The cross denotes a center of inversion and the dashed line is a mirror (a) and glide plane (b), respectively. The blue unit cell is placed around the blue sublattice, thus not containing inversion. a: symmorphic and b: nonsymmorphic variant.}
\label{fig:sketch}
\end{figure}

{\bf Figure~\ref{fig:sketch}} shows sketches of the two representative situations with a separation into different sublattices: In a first case, the space group is symmorphic, such that the space group factorizes into the translation and the point group. While the crystal has a unit cell symmetric under all point group operations, there can be sublattices of sites that sit on Wyckoff positions without inversion symmetry, see {\bf Figure~\ref{fig:sketch}a}.
Second, many space groups are nonsymmorphic, in other words some point-group operations have to be accompanied by non-trivial lattice translations in order to map the crystal lattice onto itself. Put differently, there is no choice of unit cell, such that the unit cell is invariant under all point-group operations. In this case, no Wyckoff position can have the full point group as site symmetry group. An example for a nonsymmorphic  situation is shown in \textbf{Figure \ref{fig:sketch}b}.

For concreteness, we discuss in this review systems with point group $\mathcal{G}$ involving two sublattices $A$ and $B$, denoted by orange and blue in the examples of {\bf Figure~\ref{fig:sketch}}, respectively. They are subject to a sublattice point group $\mathcal{G}_A=\mathcal{G}_B \subset \mathcal{G}$ and we assume that the sublattices lack inversion, $\mathcal{I} \notin \mathcal{G}_A$.  Note, again, that for a symmorphic crystal structure, the operations in $\mathcal{G}\setminus\mathcal{G}_A$ are pure point group operations, while for nonsymmorphic crystal structures, these operations contain non-trivial lattice translations.

\subsection{Examples of Locally Noncentrosymmetric Superconductors}

\begin{figure}[tt]
    \includegraphics{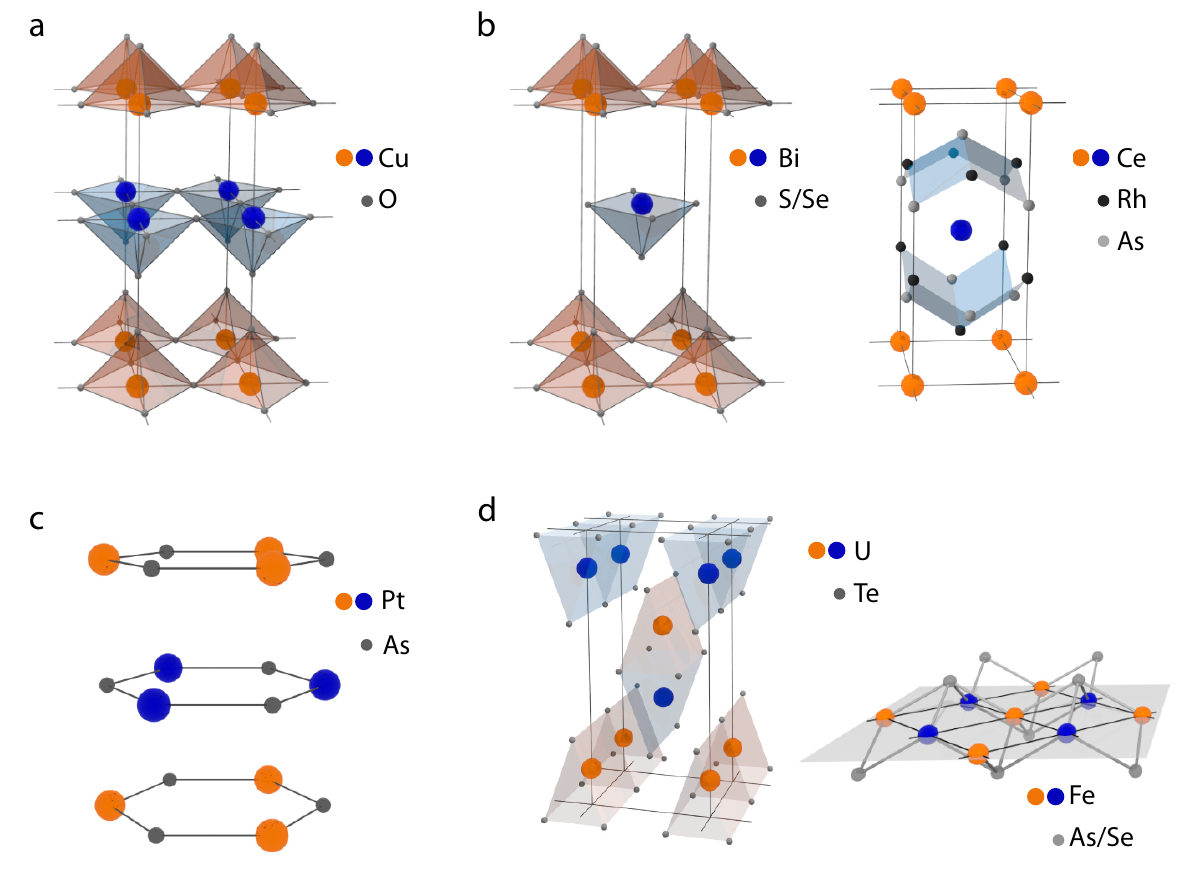}
    \caption{Illustrations of locally non-centrosymmetric crystal structures focusing on the two sublattices in blue and orange. {\bf a}  and {\bf b}: symmorphic and nonsymmorphic examples of layers lacking a mirror $M_z$, such as realized in Bi$_2$Sr$_2$CaCu$_2$O$_{8+\delta}$ or LaO$_{0.5}$Fe$_{0.5}$BiSe$_2$ and CeRh$_2$As$_2$, respectively. {\bf c}: SrPtAs (omitting Sr spacer layers) with sublattices that each lack $C_2$. {\bf d}: strongly-coupled sublattices, such as in UTe$_2$ or in single layer Fe-based superconductors.}
    \label{fig:examples}
\end{figure}

We discuss here three main classes of crystal structures with examples in \textbf {Table \ref{tab:examples}}, which have a locally-noncentrosymmetric crystal structure, see \textbf{Figure~\ref{fig:examples}}. The first class features a layered structure, where each layer lacks the in-plane mirror symmetry $M_z$ ($z\mapsto -z$) but the layers are connected through inversion with a center between the layers, such that the sublattices A and B are the two layers per unit cell. 
A symmorphic example, illustrated in \textbf{Figure~\ref{fig:examples}a}, is the quasi-tetragonal cuprate Bi$_2$Sr$_2$CaCu$_2$O$_{8+\delta}$ (Bi2212)~\cite{miles:1998,Gotlieb_2018,Atkinson_2020,Lu_2021}.
This class further includes the tetragonal CeRh$_2$As$_2$~\cite{madar:1987} and the family of BiS$_{2-n}$Se$_{n}$ superconductors, such as LaO$_{0.5}$Fe$_{0.5}$BiSe$_2$~\cite{hoshi:2022}, all showing intriguing magnetic properties. They crystalize in the nonsymmorphic P4/nmm space group (\#129) with generating point group $D_{4h}$, see \textbf{Figure~\ref{fig:examples}b} for examples. However, an individual layer is invariant only under transformations in $C_{4v}$.
An example of a trigonal crystal structure is Cu$_x$Bi$_2$Se$_3$, with crystal structure R$\overline{3}$m (\#166), which has point group $D_{3d}$, but individual layers have $C_{3v}$. Finally, artificial superlattices of heavy-fermion multilayers~\cite{shishido:2010} and bilayer Rashba systems~\cite{nakosai:2012} also belong to this class, albeit without translation symmetry in the $z$ direction. In the following, we will mostly focus on cases of this class.

\begin{table}[tt]
\tabcolsep7.5pt
\caption{Examples of locally noncentrosymmetric superconductors}
\label{tab:examples}
\begin{center}
\begin{tabular}{@{}l|c|c|c@{}}
\hline
Compound &local symmetry&point group&space group\\
\hline
CeRh$_2$As$_2$ & $C_{4v}$ & $D_{4h}$ & P4/nmm (\#129)~\cite{madar:1987}\\
LaO$_{0.55}$F$_{0.45}$BiS$_{2}$ & $C_{4v}$ & $D_{4h}$ & P4/nmm(\#129)~\cite{wu:2017}\\
Bi2212 & $C_{2v}$ & $D_{2h}$ &  Bbmb (\#66)~\cite{miles:1998}$^{\rm a}$\\
Cu$_x$Bi$_2$Se$_3$ & $C_{3v}$ & $D_{3d}$ & R$\overline{3}$m (\#166)~\cite{hor:2010}\\
\hline
SrPtAs & $D_{3h}$ & $D_{6h}$ & P$6_3$/mmc (\#194)~\cite{wenski:1986}\\
2H / 4Hb -TMDs & $D_{3h}$ & $D_{6h}$ & P$6_3$/mmc (\#194)~\cite{wilson:1969}\\
Ba$_6$Nb$_{11}$S$_{28}$ & $D_{3}$ & $D_{3d} $ & P$\overline{3}$1c (\#164)~\cite{devarakonda_2020}\\
\hline
Fe-based SC & $D_{2d}$ & $D_{4h}$ & P4/nmm (\#129)$^{\rm b}$~\cite{johrendt:2011} \\
UTe$_2$ & $C_{2v}$ & $D_{2h}$ & Immm (\#71)~\cite{Hutanu:ra5072}\\
UPt$_3$ & $D_{3h}$ & $D_{6h}$ & P6$_3$/mmc (\#194)~\cite{Joynt_2002}\\
UCoGe & $C_{1h}$ & $D_2h$ & Pnma (\#62)~\cite{canepa:1996}
\end{tabular}
\end{center}
\begin{tabnote}
    $^{\rm a}$ alternatively, the noncentrosymmetric space group Bb2b (\#37) was suggested~\cite{ivanov:2018}; $^{\rm b}$ or I4/mmm (\#139).
\end{tabnote}
\end{table}

A second class comprises crystal lattices, where the individual layers possess an in-plane mirror, but lack a $C_2$ rotation symmetry with axis perpendicular to the plane. Again, the sublattices A and B are the two layers in the unit cell. Here, famous examples are the hexagonal SrPtAs~\cite{wenski:1986, nishikubo:2011, youn:2012}, illustrated in \textbf{Figure~\ref{fig:examples}c}, and 4Hb-TaS$_2$~\cite{disalvo:1973}, both of which have been suggested to be chiral superconductors~\cite{biswas:2013, ribak:2020}. More generally, transition metal dicalchogenides (TMDs) in the 2H polytype, such as NbSe$_2$, have the same symmetry, namely P6$_3$/mmc (\#194) with point group $D_{6h}$ and layer symmetry $D_{3h}$~\cite{wilson:1969}. More recently, the layered material Ba$_6$Nb$_{11}$S$_{28}$ with point group $D_{3d}$ and layer symmetry $D_3$ has been fabricated and exhibits large critical fields \cite{devarakonda_2020}.

A third class that is worthwhile mentioning are materials with inversion-related A and B sublattices, but these sublattices are strongly coupled. While such materials do not typically reveal the physics more commonly associated with local inversion-symmetry-broken superconductivity (which will be discussed later), they do reveal interesting physical properties that are a consequence of this sublattice structure.  The first example of this type is the iron-based superconductors (Fe-based SC), where the sublattices are the A and B sublattices of the bipartition of the square lattice, each individually lacking inversion due to the out-of-plane positions of the As or Se atoms. A single layer of this class is illustrated in \textbf{Figure~\ref{fig:examples}d}. Depending on the chemical composition, the space group is either tetragonal symmorphic (I4/mmm - \#139) or nonsymmorphic (P4nmm - \#129). For the sublattice symmetry, the relevant symmetry is in both cases $D_{2d}$, which lacks both $C_4$ and inversion, but contains $C_2$, together leading to intricate spin-orbit-coupling terms~\cite{fischer:2013b,cvetkovic:2013}. Other examples include UTe$_2$~\cite{Ran_2019,Aoki_2022}, see \textbf{Figure~\ref{fig:examples}d}, paramagnetic UCoGe~\cite{Aoki_2019}, and UPt$_3$~\cite{Joynt_2002}. In all these materials the U atoms do not sit on inversion centers, a feature that might be related to the appearance of odd-parity superconductivity \cite{Anderson_1985}.

\subsection{Normal State and Microscopic Hamiltonian}

Before discussing superconductivity, it is worthwhile to address the consequences of local inversion-symmetry breaking for the normal state of electrons. For this purpose, we formulate a tight-binding Hamiltonian, which can be written in momentum space as
\begin{equation}
    \mathcal{H} = \sum_{\Vk} \vec{C}_{\Vk}^{\dag} H_{\Vk} \vec{C}_{\Vk}^{\phantom{\dag}}
\end{equation}
with the momentum $\Vk$. The creation and annihilation operators $\vec{C}_{\Vk}^{\dag}$ and  $\vec{C}_{\Vk}^{\phantom{\dag}}$ are spinors containing all local degrees of freedom such as sublattice, spin, or atomic orbitals and $H_{\Vk}$, correspondingly, has matrix form.
A generic Hamiltonian can be constructed noting that it has to transform as a scalar under all point group operations of the crystal lattice $g \in \mathcal{G}$---the pure translations have been taken care of by working in momentum space. Concretely, we require for $H_{\Vk}$ that
\begin{equation}
    U_{\Vk}(g) H_{\Vk} U_{\Vk}(g)^\dag = H_{D(g)\Vk},
\end{equation}
where $U_{\Vk}(g)$ is the representation of $g$ on the space of local degrees of freedom and $D(g)$ denotes the corresponding transformation of the (momentum) vectors. Finally, the Hamiltonian is constrained by time-reversal symmetry $\mathcal{T}$.

\begin{figure}
	\includegraphics{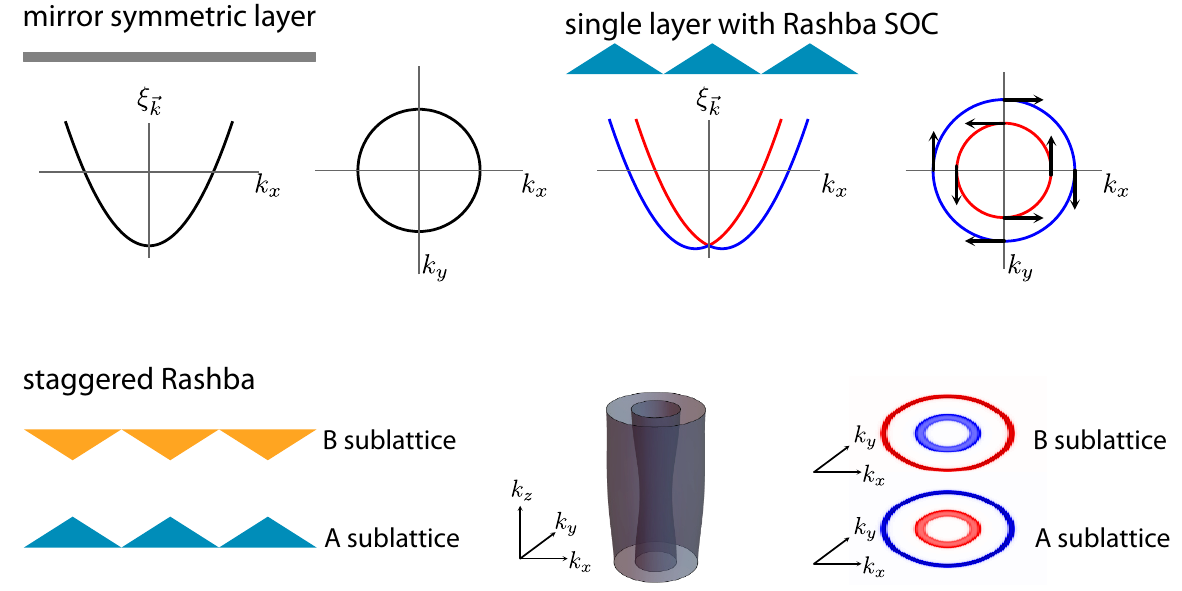}
	\caption{Comparison of inversion-symmetric layer, a layer with Rashba spin-orbit coupling, and a staggered-Rashba system. For the single layer with Rashba ASOC, the spin winds around the Fermi surface as denoted by the blue and red band. While the two bands shown for the locally-noncentrosymmetric system are two-fold degenerate, the spin and momentum are locked as in the Rasbha system when considering the layer-resolved spectral density, see bottom right.}
	\label{fig:staggered}
\end{figure}

We can construct the Hamiltonian by working in the sublattice basis $\{\vec{C}^A_{\Vk}, \vec{C}^B_{\Vk}\}$ and introducing Pauli matrices $\tau^i$ with $i=1,2,3$ and identity $\tau^0$ acting on this space. In this basis, $H_{\Vk}$ has a tensor-product structure, 
\begin{equation}
    H_{\Vk} = \sum_\nu H^\nu_{\Vk}\otimes\tau^\nu
    \label{eq:Hsp}
\end{equation}
with $\nu = 0,1,2,3$. To gain an understanding of the physical significance of the $\tau$ matrices, let us initially consider the Hamiltonian in a single sublattice, for example sublattice $A$ and, for simplicity, restrict to one atomic orbital. The generalization to more degrees of freedom is straight forward. As an illustrative example, we consider the case of a stack of square lattices, such that $\mathcal{G} = D_{4h}$ and $\mathcal{G}_A = C_{4v}$. We separate this single-sublattice Hamiltonian into two parts, which we denote by $H^\mathcal{G}_{\Vk}$ and $H^{\mathcal{G}_A}_{\Vk}$. The first one contains all the terms generically allowed in a system with point group $\mathcal{G}$, thus containing all the (spin-independent) hopping terms of the square lattice leading to a dispersion $\epsilon_{\Vk}$, while the second one contains all additional terms that are allowed for point group $\mathcal{G}_A$. Importantly, this second term contains an anti-symmetric spin-orbit coupling (ASOC) of Rashba type $ \vec{\lambda}_{\Vk} \cdot \vec{\sigma}$ with  $\vec{\lambda}_{\Vk}=-\vec{\lambda}_{-\Vk}$. This term lifts the spin degeneracy of the bands with two split bands $\xi_{\Vk, \pm} = \epsilon_{\Vk} \pm |\vec{\lambda}_{\Vk}|$ in the simplest case. The standard Rashba-type ASOC leads to the well-known spin splitting pattern as depicted in the top row of {\bf Figure~\ref{fig:staggered}} and plays a central role in locally noncentrosymmetric superconductors.

Intuitively, ASOC due to the local lack of inversion symmetry can be attributed to a polar environment within the sublattice. This environment yields $  \vec{\lambda}_{\Vk} \cdot \vec{\sigma}$ through the relativistic magnetic field experience by a moving electron in the corresponding electric field, $ \vec{B} \cdot \vec{\sigma} = -(\vec{v}_{\Vk} / c \times \vec{E}) \cdot \vec{\sigma} $, where $ \vec{v}_{\Vk}$ is the electron velocity.
The lack of an in-plane mirror $M_z$ on a square lattice, for instance, leads to the generic form  $\vec{\lambda}_{\Vk} = \lambda(-\sin k_y, \sin k_x, 0) \perp \hat{z} $, while  $\vec{\lambda}_{\Vk} \parallel \hat{z}$ appears for missing $C_2$ symmetry as in SrPtAs or 2H-TMDs.
Note beyond this symmetry based intuitive picture, we can also construct $H_{\Vk}^{\mathcal{G}_A}$ straightforwardly within a tight-binding picture taking the appropriate atomic spin-orbit coupling, hopping integrals, and orbital hybridizations into account~\cite{fischer:2010}.

Including both sublattices, we find that in Equation~\ref{eq:Hsp} $H_{\Vk}^0 = H^\mathcal{G}_{\Vk}$ with all terms scalar under $ \mathcal{G} $, while  $H^{\mathcal{G}_A}_{\Vk}$ changes sign between the two sublattices and is associated with $\tau^3$, in other words $H_{\Vk}^3 = H^{\mathcal{G}_A}_{\Vk}$ having odd parity. This latter term includes the ASOC. The remaining two components are intersublattice coupling terms and typically do not include any relevant ASOC. The Hamiltonian that connects the two sublattices symmetrically, $H_{\Vk}^1$, has to be a scalar with respect to $\mathcal{G}$. The Hamiltonian $H_{\Vk}^2$  connecting the sublattices is a scalar in $\mathcal{G}_A$, but not in $\mathcal{G}$. As will become apparent later, the consequences of local inversion-symmetry breaking are strongest when the energy scales of the intrasublattice term $H_{\Vk}^3$ are larger than those of the intersublattice terms $H_{\Vk}^1$ and $H_{\Vk}^2$.

From a symmetry perspective, the tensor-product structure of the single-particle Hamiltonian in Equation~\ref{eq:Hsp} allows for a classification by means of irreducible representations of $ {\cal G} $ as the decomposition of product representations $R\otimes R'$ with $R'$ the irreducible representation corresponding to $\tau^a$~\cite{fischer:2011b}. For the sublattice part, 
it is easy to see that $ \tau^0$ and $\tau^1$ do not change under an interchange of the sublattices, such that they belong to the trivial irreducible representation $ R' = A_{1g} $ 
of $ \mathcal{G} $. Terms with $\tau^2$ and $\tau^3$, however, change sign under the interchange of sublattices and thus, belong to an irreducible representation $ \Gamma' $ specific to $ {\cal G} $ and the sublattice structure. In particular, $\Gamma'$ is the non-trivial irreducible representation of $\mathcal{G}$ that transforms trivially under all transformations in $\mathcal{G}_A$. For the example of the staggered-Rashba system with $\mathcal{G}=D_{4h}$ and $\mathcal{G}_A = C_{4v}$, we find $\Gamma' = A_{2u}$.

The two-sublattice system has both global inversion $\mathcal{I}$ and TRS $\mathcal{T}$. The combined symmetry $\mathcal{T}\mathcal{I}$ is an antiunitary symmetry that satisfies $(\mathcal{TI})^2=-1$, implying that at each momentum $\Vk$, there is a two-fold Kramer's degeneracy which is often labelled as pseudospin. For a simple tight-binding model with $H_{\Vk}^1 = \epsilon_{\Vk}^1$ and $H_{\Vk}^2 = \epsilon_{\Vk}^2$, this two-fold degenerate dispersion reads $\xi_{\Vk, \pm} = \epsilon_{\Vk} \pm \sqrt{(\epsilon_{\Vk}^1)^2+ (\epsilon_{\Vk}^2)^2+ |\lambda_{\Vk}|^2}$. 
In general, the pseudospin does not transform as usual spin-1/2  under rotations. However, it is often possible to find eigenstates of the Bloch Hamiltonian  $|\psi_{\Vk,\alpha}\rangle$ that transform under a  rotation $g$ as $|\psi_{\Vk,\alpha}\rangle\rightarrow U_{\alpha,\beta}(g)|\psi_{D(g)\Vk,\beta}\rangle$ with $U(g)$ a usual $SU(2)$ spin rotation matrix. We call such a pseudospin basis a manifestly covariant Bloch basis (MCBB) \cite{Fu_2015}. Note that when a MCBB exists, the states are generally not pure spin states, but also include other degrees of freedom, such as orbital angular momenta or sublattice degrees of freedom.
Thus, while the bands are (pseudospin) degenerate here, there can still be a `hidden local spin-momentum locking' analogous to a Rashba system~\cite{wu:2017, ivanov:2018, gotlieb:2018}, see {\bf Figure~\ref{fig:staggered}}. Concretely, the sublattice index plays a key role in defining the pseudospin degeneracy: When  $\epsilon_{\Vk}^1=\epsilon_{\Vk}^2=0$, the pseudospin-degenerate partners at $\Vk$ are the opposite spins in the two sublattices.

\subsection{Order-Parameter Classification}
To classify the superconducting order parameters with respect to the point group of the lattice~\cite{sigrist:1991}, we work with a Bogolyubov--de Gennes (BdG) Hamiltonian
\begin{equation}
    H_{\Vk}^{\rm BdG} = \begin{pmatrix} H_{\Vk} & \Delta_{\Vk} \\ \Delta_{\Vk}^\dag & - H_{-\Vk}^* \end{pmatrix},
\end{equation}
such that the transformation properties of the order parameter are given by $U_{\Vk}(g) \Delta_{\Vk} U_{-\Vk}(g)^T = \chi_g \Delta_{D(g)\Vk}$, with $\chi_g \in U(1)$ the respective eigenvalue of the order parameter.
We can use the sublattice basis introduced in the previous section to classify gap functions of the form~\cite{fischer:2011b}
\begin{equation}
  \Delta_{ss'}^{\alpha\alpha'}(\Vk) = \sum_\nu \{\psi_\nu(\Vk)(i\sigma^y) + [\vec{d}_\nu(\Vk)\cdot\vec{\sigma}](i\sigma^y)\}_{ss'}\otimes\tau^\nu_{\alpha\alpha'},
  \label{eq:generalgap}
\end{equation}
with the spin-singlet component $\psi_\nu(\Vk)$ and spin-triplet component $\vec{d}_\nu(\Vk)$ classified according to the irreducible representations of $\mathcal{G}$. 
For the full classification of the gap functions, we use the same scheme as for the Hamiltonian, namely that they transform as $R\otimes R'$ with $R'$ the irreducible representation of $\tau^\nu$, either $A_{1g}$ or $\Gamma'$. Importantly, since the crystal lattice possesses global inversion symmetry, the superconducting order parameters can be classified as even and odd under inversion.
Finally, note that the Pauli principle has to be satisfied, requiring that $\nu=0,1,3$ (``triplet in sublattice index'') for an even spin-singlet, $\psi(-\Vk) = \psi(\Vk)$, and odd spin-triplet, $\vec{d}(-\Vk) = - \vec{d}(\Vk)$, function, whereas $\nu=2$  (``singlet in sublattice index'') for odd spin-singlet and even spin-triplet order parameters.  

It is instructive to again look at a single sublattice first. For the individual sublattice, the lack of a center of inversion leads to a mixing of even and odd pairing channels, such that a general gap function defined on sublattice $A$ reads $\Delta^A_{ss'}(\Vk) = \{\psi(\Vk)(i\sigma^y) + [\vec{d}(\Vk)\cdot\vec{\sigma}](i\sigma^y)\}_{ss'}$,
where the direction of the spin-triplet component $\vec{d}_{\Vk} \parallel \hat{\lambda}_{\Vk}$ is fixed by the spin-orbit field~\cite{gorkov:2001, smidman:2017}. From a symmetry point of view, this mixing derives from the sublattice point group $\mathcal{G}_A$. We can construct even (e) and odd (o) order parameters, respectively, by changing either the sign of the singlet or triplet component when going to sublattice B. This leads to~\cite{fischer:2011b, sigrist:2014}
\begin{eqnarray}
    \Delta^{\rm even}(\Vk) &=& \psi(\Vk)(i\sigma^y)\otimes \tau^0 + [\vec{d}(\Vk)\cdot\vec{\sigma}](i\sigma^y)\otimes \tau^3,\label{eq:evengap}\\
    \Delta^{\rm odd}(\Vk) &=& \psi(\Vk)(i\sigma^y)\otimes \tau^3 + [\vec{d}(\Vk)\cdot\vec{\sigma}](i\sigma^y)\otimes \tau^0\label{eq:oddgap}.
\end{eqnarray}

Note that similar to the situation of global inversion-symmetry breaking, the direction of the $d$ vector for intrasublattice pairing is fixed, although the energy required to rotate the d vector decreases with increasing intersublattice coupling ~\cite{fischer:2011b}. Furthermore, for the case of locally noncentrosymmetric systems, additional order parameters corresponding to intersublattice pairing  (that is $\tau^2$ and $\tau^1$ pairing) appear~\cite{fischer:2011b} and have been discussed in the context of UPt$_3$~\cite{Yanase_UPt3_2016, yanaseUPt32017} and UTe$_2$~\cite{Shishidou_2021}. However, when the intersublattice coupling is weak, the $\tau_0$ and $\tau_3$ order parameters are likely to be more relevant.

\subsection{Ginzburg-Landau Free Energy}

Within the Ginzburg-Landau formalism, the structure of the even and odd order parameters in Equations~\ref{eq:evengap} and \ref{eq:oddgap} can be accounted for by introducing even- and odd-parity order-parameter components on each sublattice. In the following, we again focus on the example of layers with broken inversion symmetry and denote by $\eta_{l,j}(\vr)$ the locally even ($l={\rm e}$) and odd ($l={\rm o}$) components in layer $j$ and $\vr$ is the in-plane coordinate, such that the general order parameter is written as $\Psi(\vr, j) = (\eta_{{\rm e}, j}(\vr), \eta_{{\rm o}, j}(\vr))$.
The free-energy density for a single layer $j$ has the form 
\begin{equation}
    f^{(j)}[\Psi(\vr, j)] = f_{\rm e}^{(j)}[\eta_{{\rm e},j}(\vr)] + f_{\rm o}^{(j)}[\eta_{{\rm o},j}(\vr)] + f_{\rm eo}^{(j)}[\eta_{{\rm e},j}(\vr), \eta_{{\rm o},j}(\vr)],
    \label{eq:single-layer}
\end{equation}
where, neglecting gradient terms and external fields for now,  $f^{(j)}_l[\eta_{l,j}(\vr)] = a_l(T) |\eta_{l, j}(\vr)|^2 + b_l/2 |\eta_{l, j}(\vr)|^4$. Furthermore, there is a coupling of the even and odd components due to the inversion-symmetry breaking, $f_{\rm eo}^{(j)} = \epsilon_j / 2 (\eta_{{\rm e},j}(\vr) \eta_{{\rm o},j}(\vr)^* + c.c.)$, with $\epsilon_j = (-1)^j \epsilon$ capturing the staggered nature of the inversion-symmetry breaking. 
Depending on which order parameter has the higher critical temperature, $a_{l}(T) = a_{l,0} (T-T_{c,l})$, the full system chooses one of the two solutions, an even and an odd one with $\Psi_{\rm even}(\vr, j) = (\eta_{{\rm e}}, (-1)^j \eta_{{\rm o}})$ and $\Psi_{\rm odd}(\vr, j) = ((-1)^j \eta_{{\rm e}}, \eta_{{\rm o}})$. The subleading order changes sign from sublattice to sublattice as expected. This sign-change, however, costs energy which we can introduce as a Josephson coupling between the sublattices,
\begin{equation}
    f^{(j,j+1)}_l = J |\eta_{l,j+1}(\vr) - \eta_{l,j}(\vr)|^2.
    \label{eq:Jcoupling}
\end{equation}
For simplicity, we parametrize the Josephson coupling by a single parameter $J$, but in general the coupling strength can depend on the respective order-parameter component. 
In the Ginzburg-Landau formulation, $\epsilon / J$ thus takes the role of the dimensionless parameter that measures the importance of noncentrosymmetricity.

\section{PHYSICAL CONSEQUENCES}

We now discuss the physical consequences of local inversion-symmetry breaking on superconductivity. In Sections~\ref{sec:response} to \ref{sec:topology}, we begin by addressing how the physics of global inversion-symmetry breaking, enumerated in the Introduction, is changed when inversion symmetry is restored globally. In Section~\ref{sec:finish}, we further discuss new phenomena in locally inversion-symmetry-broken superconductors.

\subsection{Magnetic Response}\label{sec:response}

A striking property seen in noncentrosymmetric superconductors such as CeIrSi$_3$ and CeRhSi$_3$ is the enormous upper critical field for field directions perpendicular to the basal plane, which is not a mirror plane, exceeding the paramagnetic limit by far~\cite{kimura:2007, settai:2008}.   It is thus natural to ask how much of this physics survives in the case of locally broken inversion symmetry. To answer this question, we begin with examining the spin susceptibility.

\subsubsection{Spin Susceptibilities}

In noncentrosymmetric materials, the Rashba-like ASOC directly affects the structure of the spin susceptibility in the normal state: While the susceptibility in a system without spin-orbit coupling is given by the so-called Pauli susceptibility, which is an intraband response and proportional to the density of states at the Fermi level, $\rho(\epsilon_{\rm F})$, the ASOC will introduce an interband, or so-called van Vleck contribution for fields perpendicular to $\vec{\lambda}_{\Vk}$. The diagonal susceptibility for fields along the normal vector $\hat{n}$ then reads
\begin{equation}
    \chi_{\hat{n}} = \sum_{\Vk}\sum_{\alpha = \pm} (\hat{\lambda}_{\Vk}\cdot \hat{n})^2\chi_{\rm P}(\Vk, \alpha)+ \sum_{\Vk} (\hat{\lambda}_{\Vk}\times \hat{n}_i)^2 \chi_{\rm vV}(\Vk)
\end{equation}
with $\hat{\lambda}_{\Vk} = \vec{\lambda}_{\Vk} / |\vec{\lambda}_{\Vk}|$ the unit vector in the direction of the spin-orbit field; the Pauli contribution $\chi_{\rm P}(\Vk, \alpha) \propto\partial_{\xi}n_{\rm F}(\xi_{\Vk, \alpha})=S^\alpha_{\Vk}$, given by the spectral density of band $\alpha$; the van Vleck contribution $\chi_{\rm vV}(\Vk) \propto (n_{\rm F} (\xi_{\Vk, +}) - n_{\rm F}(\xi_{\Vk, -}))/(\xi_{\Vk, +} - \xi_{\Vk, -})$; and the Fermi distribution function $n_{\rm F}(\xi)$. As a consequence, a single layer with Rashba ASOC has a susceptibility that is purely van Vleck for out-of-plane fields, while the (1H) TMDs have a pure van Vleck susceptibility for in-plane fields. With the van Vleck contribution not depending on states at the Fermi level, the additional structure of the susceptibility is crucial in the superconducting state, where the Fermi surface is gapped out. In particular, when the spin-orbit energy scale is much bigger than the superconducting energy scale, $\chi_{\rm vV}(\Vk)$ is unaffected by superconductivity, leading to a residual non-zero spin susceptibility in the superconducting state---even for primarily spin-singlet superconductors.

To elucidate the fate of the residual spin susceptibility in the superconducting state when inversion symmetry is restored globally, a calculation based on the sublattice Hamiltonian of Equation~\ref{eq:Hsp} yields for fields along $\hat{n}$ for the simplest case of a single-orbital model
\begin{equation}
    \chi_{\hat{n}}=  \sum_{\Vk}\sum_{\alpha = \pm} [(\tilde{\epsilon}_{\Vk})^2 + (\tilde{\lambda}_{\Vk}\cdot \hat{n})^2]\chi_{\rm P}(\Vk, \alpha) + \sum_{\Vk} (\tilde{\lambda}_{\Vk}\times \hat{n})^2 \chi_{\rm vV}(\Vk).
    \label{eq:fullsusc}
\end{equation}
Here, $(\tilde{\epsilon}_{\Vk})^2 = ((\epsilon^1_{\Vk})^2 + (\epsilon^2_{\Vk})^2) /  ((\epsilon_{\Vk}^1)^2 + (\epsilon_{\Vk}^2)^2 + |\lambda_{\Vk}|^2)$ and $\tilde{\lambda}_{\Vk} = \lambda_{\Vk} / \sqrt{(\epsilon_{\Vk}^1)^2 + (\epsilon_{\Vk}^2)^2 + |\lambda_{\Vk}|^2}$. There is still a van Vleck contribution, although its relative importance is reduced by the intersublattice coupling.

Consequently, in the superconducting state the residual susceptibility, again given by the van Vleck contribution,  survives even for a conventional spin-singlet pairing state but is suppressed by the intersublattice coupling. A simple understanding of this result follows from projecting a Zeeman magnetic field $\sum_i H_i\sigma_i$ onto the pseudospin basis that describes a single band. This projection between spin $\sigma_i$ and pseudospin $s_i$ takes the general form 
\begin{equation}
\sigma_{\hat{n}}\rightarrow \sum_{i=\{x,y,z\}}\gamma^{\hat{n},i}_{\Vk}s_i
\end{equation}
and the field direction experienced by the pseudospin is not generally the same as the applied field direction of the Zeeman field.
The Zeeman field experienced along direction $\hat{n}$, which is given by
\begin{equation}
\sum_i(\gamma^{\hat{n},i}_{\Vk})^2=\frac{(\epsilon_{\Vk}^1)^2+ (\epsilon_{\Vk}^2)^2+(\hat{n}\cdot\vec{\lambda}_{\Vk})^2}{(\epsilon_{\Vk}^1)^2+ (\epsilon_{\Vk}^2)^2+ |\lambda_{\Vk}|^2},
\label{Zsquare}
\end{equation}
is thus in general less than the applied magnetic field and corresponds to the prefactor of the Pauli susceptibility in Equation~\ref{eq:fullsusc}.

In addition to the van Vleck contribution, for odd-parity superconductors there is another well-known source for a residual susceptibility. In particular, odd-parity superconductors are typically described by a vector $\Vd(\Vk)$, see Equation~\ref{eq:generalgap}, whose components transform as a vector under rotations~\footnote{This property implicitly assumes that a MCBB pseudospin basis can be found}. If this $d$ vector is orthogonal to the applied magnetic field for all $\Vk$, the field can spin polarize all Cooper pairs and the spin susceptibility in the superconducting state is unchanged from that of the normal state. For the case of a staggered Rashba system with $\hat{\lambda}_{\Vk}\perp \hat{z}$, this is the case for out-of-plane fields for intrasublattice spin-triplet states with $\vec{d}_{\Vk} \parallel \hat{\lambda}_{\Vk}$~\cite{skurativska:2021b}.

\subsubsection{Large Pauli limiting fields}

Upper critical fields exceeding the expected Pauli limiting fields have indeed been observed in several of the systems introduced in Section~\ref{sec:basics}, most strikingly in bilayer NbSe$_2$ realizing a 2H polytype~\cite{barrera:2018}. Furthermore, such an effect has been observed for in-plane fields in LaO$_{0.5}$F$_{0.5}$BiS$_2$~\cite{chan:2018}, LaO$_{0.48}$F$_{0.52}$BiSe$_2$~\cite{shao:2014}, and LaO$_{0.5}$F$_{0.5}$BiSSe~\cite{kase:2017} and most recently in LaO$_{0.5}$F$_{0.5}$BiS$_{2-x}$Se$_{x}$ $(x=0.22, 0.69)$~\cite{hoshi:2022} all with P4/nmm space group, while out-of-plane fields seem to be orbitally limited. Similarly, the high upper critical field in CeCoIn$_5$ heterostructures~\cite{mizukami:2011,Shimozawa_2016} has been suggested to come from the locally noncentrosymmetric structure~\cite{maruyama:2012a}.
To better understand these findings, we first review the relative importance of the two ways a magnetic field couples to and destroys superconductivity: paramagnetic and orbital depairing.

A magnetic field coupling to the electrons' spins through a Zeeman term lifts the degeneracy between states with opposite momentum and spin and hence, suppresses superconductivity. Put differently, the superconducting state cannot be polarized and as such cannot optimize the magnetic energy. Once this magnetic energy is of the order of the condensation energy, the system gains energy entering the normal state, thus leading to the paramagnetic limiting field $H_{\rm p}(T) \sim k_{\rm B} T_{\rm c} / \sqrt{\chi^{\rm n} - \chi^{\rm sc}(T)}$. For a conventional superconductor with a pure Pauli susceptibility, in other words $\chi^{\rm n} = \mu_{\rm B}^2 \rho(\epsilon_{\rm F})$ and $\chi^{\rm sc}(T=0) = 0$, this yields $H_{\rm p} \sim \Delta / \mu_{\rm B}$ known as the Clogston-Chandrasekhar limit~\cite{clogston:1962, chandrasekhar:1962}. A second way of depairing Cooper pairs in a superconductor is through the orbital coupling of the magnetic field. Here, the depairing happens when vortices start to overlap at the orbital upper critical field $H_{\rm c2}^{\rm orb} \sim \phi_0 / \xi_0^2$ with $\phi_0 = h / (2e)$ the flux quantum in the superconductor and $\xi_0$ the coherence length at $T=0$. The relative importance of these two ways of depairing a superconductor is captured by the Maki parameter $\alpha_{\rm M} = \sqrt{2}H_{\rm c2} (0) / H_{\rm p} (0)$~\cite{maki:1966}. Concretely, the renormalization of the spin susceptibility and hence the enhancement of the paramagnetic depairing field only becomes relevant for pair breaking for large $\alpha_{\rm M}$. 

While large orbital upper critical fields follow from very short coherence lengths not uncommon for unconventional superconductors, spin-orbit coupling has a profound effect on the paramagnetic critical field through its change to the spin susceptibility both in the normal and the superconducting state as discussed above~\cite{maruyama:2012a, youn:2012, skurativska:2021b}.
If the intersublattice coupling terms $\epsilon_{\Vk}^1=\epsilon_{\Vk}^2=0$ and the field can be chosen to be orthogonal to the ASOC, that is $\vec{H}\cdot\vec{\lambda}_{\Vk}=0$, the Pauli limiting field diverges. For the example of a pure Rashba ASOC, a Zeeman field applied along the $c$-axis will have no effect---this remains correct as long as the Zeeman energy scale is much less than any interband separation energy.

\subsection{Local parity mixing and field driven even- to odd-parity transitions}\label{sec:parity-transition}

When the two inversion-symmetry related sublattices are weakly coupled, it is natural to assign a superconducting order parameter to each sublattice.
Assuming only spin-singlet even-parity pairing within the planes, the even- and odd-parity states of Equations~\ref{eq:evengap} and \ref{eq:oddgap} should be nearly degenerate. This near degeneracy allows us to use symmetry-breaking fields, for example magnetic fields, to control which pairing state appears.  This possibility  was first discussed in the context of two weakly coupled CeCoIn$_5$ layers~\cite{yoshida:2012}, for which  a natural pairing symmetry for each layer is a $d$-wave spin-singlet state, and it was pointed out that a $z$-axis field drives a transition to an odd-parity `pair density wave' state. 

Remarkably, in CeRh$_2$As$_2$ a field-induced phase transition that strongly resembles that predicted for bilayer CeCoIn$_5$ has been observed~\cite{khim:2021}. In particular, the upper critical field for fields directed along the $z$ axis (perpendicular to the layers) extrapolates to $\sim15$ T at zero temperature, far beyond the paramagnetic limiting field $H_{\rm p} \sim 0.5$ T for a critical temperature of $T_{\rm c} \approx 0.26$ K.
Moreover, the upper critical field shows a pronounced kink for a temperature $T$ roughly half of $T_{\rm c}$ and several thermodynamic quantities indicate a first-order transition within the superconducting phase. These observations strongly hint towards a switch in the order-parameter symmetry upon increasing magnetic field~\cite{khim:2021, moeckli:2018, schertenleib:2021, nogaki:2021}. In the following, we initially examine the field-induced phase transition microscopically and follow with a phenomenological description that reveals additional physics is possible.

\subsubsection{Microscopic description of field-induced even- to odd-parity transitions}

For an intuitive microscopic picture of the field-induced transition, it helps to rewrite the nearly degenerate even- and odd-parity states in the pseudospin basis. Starting in the sublattice basis, the even-parity state takes the form $\Delta^{\rm even} = \Delta_{\rm e}(i\sigma^y)\otimes \tau^0$, while the odd-parity state takes the form $\Delta^{\rm odd}.= \Delta_{\rm o}(i\sigma^y)\otimes \tau^3$. Including a staggered Rashba ASOC $\vec{\lambda}_{\Vk}\cdot\vec{\sigma}\otimes\tau^3$, with  $\vec{\lambda}_{\Vk} = \lambda(-\sin k_y, \sin k_x, 0) $, we find
\begin{eqnarray}
\Delta^{\rm even}&&\rightarrow \Delta_{\rm e} is^y\\
\Delta^{\rm odd}&&\rightarrow \frac{\lambda(\sin k_x s^y-\sin k_y s^x)}{\sqrt{ (\epsilon^1_{\Vk})^2+(\epsilon^2_{\Vk})^2+\lambda^2(\sin^2 k_x+\sin^2k_y)}}\Delta_{\rm o}is^y.
\end{eqnarray}
These expressions immediately reveal the physics of the phase diagram for fields along the $z$ axis: Since the $d$ vector for $\Delta_{\rm o}$ is purely in plane, it is not suppressed by a field along the $z$ axis, while $\Delta_{\rm e}$ is, such that the field can enable the transition. In addition, notice that the magnitude of the prefactor multiplying $\Delta_{\rm o}$ in the pseudospin basis is generally less than one, implying the $T_{\rm c}$ for the odd-parity state will be suppressed relative to that of the even-parity state. Furthermore, notice that this magnitude becomes one when $\epsilon^1_{\Vk}=\epsilon^2_{\Vk}=0$. In this limit, the $T_{\rm c}$ for the even- and odd-parity cases become degenerate, as intuitively expected in the limit in which the two sublattices are uncoupled. This illustrates again the importance of having the spin-orbit energy scales larger than the intersublattice coupling to reveal the physics of local inversion-symmetry breaking.

\subsubsection{Ginzburg-Landau description of field-induced even- to odd-parity transitions}
\begin{figure}[tt]
\includegraphics{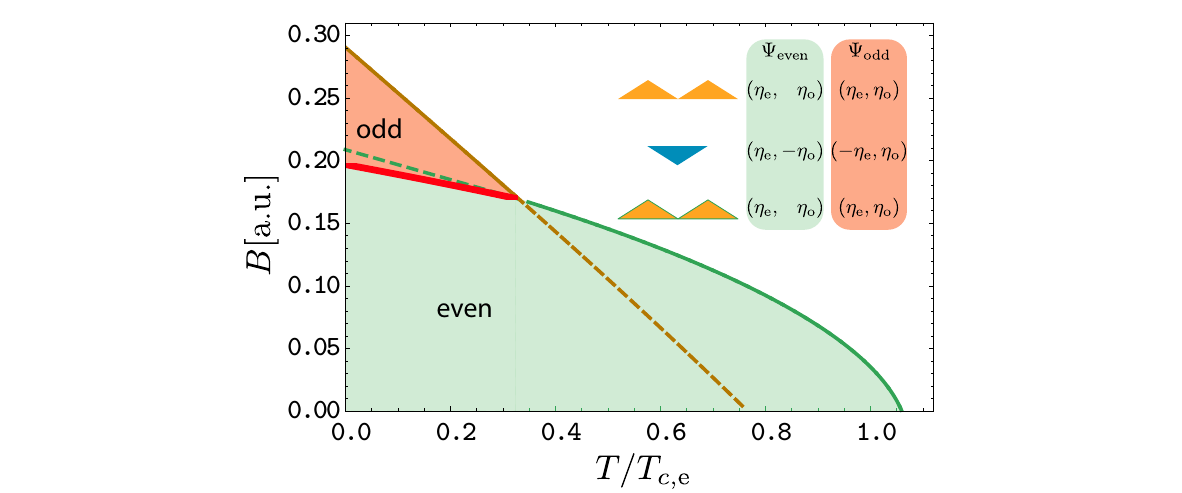}
\caption{Phase diagram for $\vec{B}\parallel\hat{z}$ with even-parity low-field and odd-parity high-field phases adapted from~\cite{schertenleib:2021}. The red line denotes the first-order transition line found using a circular-cell method. The inset shows the structure of the globally even and odd solutions.}
\label{fig:hz}
\end{figure}
We can phenomenologically study the $B-T$ phase diagram by introducing the vector potential $\vec{A}$ and associated magnetic field $\vec{B} = \vec{\nabla}\times \vec{A}$ to the free-energy density introduced in Equations~\ref{eq:single-layer} and \ref{eq:Jcoupling}. The magnetic field couples, on the one hand, directly via a Pauli-limiting term, here only to the even spin-singlet component. On the other hand, the vector potential enters through minimal coupling to the orbital part, such that the full free-energy density for a single layer reads
\begin{equation}
    f^{(j)}[\Psi(\vr, j), \VA] = f^{(j)}[\Psi(\vr, j)] + \frac{\VB^2}{8\pi}  + \frac{d \chi^{\rm n}}{2}\VB^2 |\eta_{{\rm e}, j}(\vr)|^2 + \sum_l \frac{1}{2m_l} |\boldsymbol{D}_{\parallel} \eta_{l, j} (\vr) |^2.
\end{equation}
Here, $\boldsymbol{D}_{\parallel} = [-i\hbar \boldsymbol{ \nabla}+2e{\bf A}/c]_{\parallel}$ is the in-plane component of the covariant derivative, $\chi^{\rm n}$ the normal-state spin susceptibility, and $d$ a coupling constant. While only the even in-plane component is directly paramagnetically limited, both solutions are limited through the parity mixing within the plane.
Finally, for a field $\VB \parallel \hat{z}$, the magnetic field reduces the critical temperature linearly through orbital effects. The resulting phase diagram with a field-induced first-order transition is shown in {\bf Figure~\ref{fig:hz}}~\cite{schertenleib:2021}.

The first-order phase transitions can qualitatively be envisaged as the inversion-symmetric analogue of the even-odd parity mixing found in noncentrosymmetric superconductors. 
When inversion symmetry is globally broken, then even- and odd-parity states will be generically mixed and no such phase transition will occur. This explains why noncentrosymmetric CeIrSi$_3$ and CeRhSi$_3$ exhibit similar critical field anisotropies to locally broken inversion symmetric CeRh$_2$As$_2$ but do not exhibit the field-induced phase transition observed in CeRh$_2$As$_2$. 
For a generic centrosymmetric system, on the other hand, the first-order transition can split up into two second-order transitions.
In order to study the stability of the first-order transition, we consider the more generic forth-order terms in the free-energy density,
$
    f^{(j)}_4[\Psi(\vr, j)] = \sum_l\frac{b_l}{2}|\eta_{l, j}|^4 + b_2 |\eta_{{\rm e}, j}||\eta_{{\rm o}, j}| + b_3 [\eta_{{\rm e}, j}^2 (\eta_{{\rm o}, j}^*)^2 + (\eta_{{\rm e}, j}^*)^2\eta_{{\rm o}, j}^2].
$
Without the linear coupling between $\eta_{\rm e}$ and $\eta_{\rm o}$ in each layer, the system can gain energy for $b_3>0$ by entering a TRS-breaking state $(\eta_{\rm e}, \pm i \eta_{\rm o})$ right at $T_{\rm c}$. The physical reason for such a solution  is that the system can gain condensation energy by avoiding gap nodes. As a consequence, the first-order line is split up into two second-order transitions, a generic feature observed, for example, in UPt$_3$~\cite{adenwalla:1990}. For the case of a locally-noncentrosymmetric system, however, the two order parameters are coupled linearly in each layer. 
A TRS-breaking solution is still possible, where
\begin{equation}
    \Psi(\vr, j) = \left\{\begin{array}{ll} (\eta_{\rm e}, \eta_{\rm o} e^{2i\phi})  & j = 2n\\ (\eta_{\rm e}e^{-i\phi}, - \eta_{\rm o}e^{i\phi}) & j = 2n+1 \end{array}\right.
\end{equation}
with $n\in\mathbb{Z}$ and $\phi\in[0, 2\pi)$. However, such a solution can become stable only at $T^* < T_{\rm c}$ and does not remove the first-order transition.

\subsection{From helical phase to complex stripe phase}

Magnetic fields applied in plane to mirror-symmetry-broken superconductors, such as CeRhSi$_3$~\cite{kaur:2005a}, the LAO/STO interface~\cite{michaeli:2012a}, or surface superconductivity~\cite{barzykin:2002} have been suggested to produce a helical phase, similar to the finite-momentum phase suggested by Fulde and Ferrell~\cite{fulde:1964}. 
The helical phase can be understood on symmetry grounds, since the simultaneous breaking of inversion and time-reversal symmetries allows for Lifshitz invariants to appear in the Ginzburg-Landau free energy, which stabilize these finite-momentum pairing states~\cite{agterberg:2003, smidman:2017}.  Here, we address the fate of this phase when inversion symmetry is restored.

For this purpose, we work with a Ginzburg-Landau description and restrict ourselves to the case of thin layers and ignore the in-plane modulations of the vector potential. Furthermore, we consider a single order parameter $\Psi(\vr, j) = \eta_{{\rm e}, j}(\vr) = \eta_j(\vr)$, as the additional spin-triplet component considered for out-of-plane fields is now also Pauli limited and hence, not favorable. Again, we start from a single layer for an in-plane field $\vec{B}\perp \hat{z}$. In analogy to the Rashba ASOC, there is a Lifshitz invariant in the free energy~\cite{edelstein:1996}
\begin{equation}
    f^{(j)}_{\rm lif} = (-1)^j\varepsilon(\hat{z}\times \vec{B})\cdot(\eta_j^*(\vr) \boldsymbol{D}_\parallel \eta_j(\vr) + \eta_j(\vr) (\boldsymbol{D}_\parallel \eta_j(\vr))^*).
\end{equation}
For the individual layer with broken mirror symmetry, or equivalently in the limit $J\rightarrow0$, this Lifshitz invariant leads to the helical phase, with an order parameter $\eta_j(\vr)= \eta_0 \exp(i\vq_j\cdot \vr)$ and $\vq_j \propto (-1)^j \varepsilon (\hat{n}\times\vec{B})$~\cite{agterberg:2003}. Note that formally, the order parameter acquires a finite total momentum $\vq$ at any $B>0$. However, this phase factor can be gauged away in the bulk, such that only in frustrated situations, such as in combination with a centrosymmetric superconductor in a Josephson junction~\cite{kaur:2005a}, the phase results in physical consequences~\cite{aoyama:2012}. In these situations, not unlike the situation studied below, a finite magnetic field is necessary to see effects of the helical phase. 

\begin{figure}[tt]
    \includegraphics{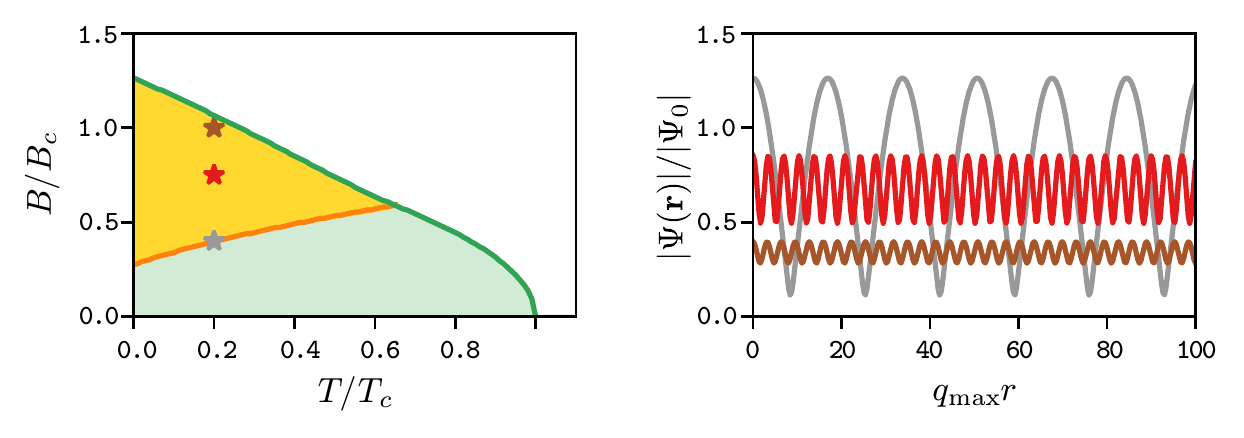}
    \caption{Phase diagram for in-plane fields with a high-field complex stripe phase found with the Ginzburg-Landau free energy in Equation~\ref{eq:helicalfree} with the paramagnetic limiting field $B_c = \sqrt{T_{\rm c} / (d\chi^n)}$. The right side shows the order parameter relative to the homogenous zero-field result $\Psi_0$ for three different magnetic fields as indicated in the phase diagram. The coordinate axis is scaled with $q_{\rm max}$, the maximal wave vector at the transition $B=B_{\rm c}^{\rm cs}$ }
    \label{fig:complex_stripe}
\end{figure}

For finite Josephson coupling $J$ between the layers, the alternating sign of $\vq_j$ leads to phase frustration, which competes with the magnetic energy gained for finite $\vq$. To see this, we make the ansatz $\eta_j(\vr) = \eta_0 (e^{i\vq_j\cdot \vr} + \delta e^{-i\vq_j\cdot \vr})$
with $\vq_j=(-1)^j \vq$ and $\eta_0$ and $\delta$ are treated as variational parameters.
After integration over the in-plane coordinate and summation over all layers, the free energy reads
\begin{multline}
    F[\eta_0, \delta] = a(T)\eta_0^2 (1+\delta^2) + d\chi^{\rm n}B^2 \eta_0^2 (1+\delta^2) + \frac{b}{2}\eta_0^2 (1 + 4 \delta^2 + \delta^4)\\
    +  2 J \eta_0^2 (1-\delta)^2 - 2 \varepsilon^2 m B^2 \eta_0^2\frac{(1-\delta^2)^2}{1+\delta^2}, \label{eq:helicalfree}
\end{multline}
where we have already used the minimizing $\vq = 2 \varepsilon m (\hat{z}\times\VB) (1-\delta^2)/(1+\delta^2)$.
From Equation~\ref{eq:helicalfree} we see explicitly that there is a competition between the inter-layer Josephson term and the ASOC term: A critical in-plane field $B_{\rm c}^{\rm cs}\sim\sqrt{J/2 m \varepsilon^2}$ is required, above which this complex stripe phase is established. 

{\bf Figure~\ref{fig:complex_stripe}} shows the phase diagram and order parameter obtained by minimizing the free energy, Equation~\ref{eq:helicalfree}, showing a complex stripe phase, which right above $B^{\rm cs}_{\rm c}$ is a spatially modulated phase akin to the Larkin-Ovchinikov phase~\cite{larkin:1965}. With increasing magnetic field, the layers become increasingly decoupled, such that $\delta \rightarrow 0$ and the spatial modulation decreases, similar to the $J=0$ case. This picture is consistent with similar results seen within a BdG approach~\cite{yoshida:2013, matsutomi:2020}
and could be realized in LaO$_{0.5}$F$_{0.5}$BiS$_{2-x}$Se$_{x}$ $(x=0.22, 0.69)$~\cite{hoshi:2022}. Finally, note that incorporation of orbital effects in a multi-layer system  may stabilize an odd-parity superconducting state similar to that discussed in Sec.~3.1.2~\cite{watanabe2015}.

\subsection{Topology and spontaneous TRS breaking}\label{sec:topology}
Fully-gapped superconducting phases can be classified according to the ten-fold way~\cite{ryu:2010}. Having explicit particle-hole symmetry in the BdG Hamiltonian, superconducting phases generically fall into classes DIII (with TRS) or D (without TRS) in the AZ classification. This classification allows for topologically non-trivial phases in three dimensions only in class DIII, with the topological nature of the superconducting state captured by a winding number $\nu$~\cite{schnyder:2008}. Class D, on the other hand, has no strong topological phase. In two dimensions, both classes allow for non-trivial phases, characterized by a $\mathbb{Z}_2$ invariant (class DIII) and a $\mathbb{Z}$-valued invariant, the Chern number, respectively.

While the topological invariants can, in principle, be calculated explicitly, for an inversion symmetric system, $\mathcal{I}\mathcal{H}_{\Vk}\mathcal{I}^{-1} = \mathcal{H}_{-\Vk}$, topological superconducting phases can also be inversion-symmetry indicated.
First discussed in the context of superconductivity in Cu-doped Bi$_2$Se$_3$ by Fu and Berg~\cite{fu:2010}, an inversion-symmetry indicator allows to calculate $\nu$ modulo $2$, in analogy to the indicator introduced for topological insulators~\cite{fu:2007b}. In particular, for an odd-parity order parameter, $(-1)^\nu = \prod_\alpha(-1)^{N_{\Gamma_\alpha}}$, where $N_{\Gamma_\alpha}$ is the number of occupied bands at the time-reversal invariant momentum (TRIM) $\Gamma_\alpha$. These are the momenta, where the little group contains inversion, such that $-\Gamma_\alpha = \Gamma_\alpha$ and the Bloch functions of the normal-state bandstructure are eigenfunctions of inversion. It then follows that an odd-parity order parameter on a bandstructure with an odd number of bands around TRIMs is necessarily topological. Similar result hold in 2D~\cite{sato2010} and can be applied to the Rashba bilayer system. Here, a chemical potential in the hybridization gap of the two layers leads to a topological phase~\cite{nakosai:2012}.

\subsubsection{Topological crystalline phases}\label{sec:TCSC}
Inversion symmetry indicators can be generalized to higher-order topological superconductivity~\cite{skurativska:2020, ono:2019, ono:2020}. While first-order topology is associated through the bulk-boundary correspondence with topological surface states, higher-order topological states are connected to topological hinge and corner states, respectively. Specifically, we can define the invariant
$	\kappa_{\rm dD} = \frac{1}{2} \sum_{\alpha} (n^+_{\Gamma_\alpha} - n^{-}_{\Gamma_\alpha})$,
where now, $n^\pm_{\Gamma_\alpha}$ counts the number of occupied bands at $\Gamma_\alpha$, which are even ($+$) and odd ($-$) under inversion, respectively. This indicator is a $\mathbb{Z}_8$ quantity in $d=3$ dimensions that in addition to first-order topology with $\kappa_{\rm 3D}$ odd, can indicate second-order ($\kappa_{\rm 3D} = \pm 2$) and third-order topology ($\kappa_{\rm 3D}=4$). 

In addition to topology protected by inversion symmetry, also a mirror symmetry such as present in {\bf Figure~\ref{fig:sketch}a} allows to protect topological superconducting phases. In this case, a mirror Chern number can be defined and non-trivial topology implies zero-energy surface states at surfaces that preserve the mirror. For multi-layer systems, it was found that for odd numbers of layers and an odd-parity order parameter, the system can indeed be topological~\cite{yoshida:2015}. Unlike the strong topology found in bilayer Rashba systems~\cite{nakosai:2012}, this crystalline topology does not require fine tuning of the chemical potential. Furthermore, a field applied perpendicular to the plane does not break the mirror symmetry, such that this phase carries over to a field-induced odd-parity phase.

The mirror symmetric systems thus fall into a $\mathbb{Z} \times \mathbb{Z}$ classification in accordance with the K-theory for noninteracting systems~\cite{shiozaki2014}. However, the classification is reduced by interactions, as shown for one-dimensional superconductors~\cite{Z_to_Zn_Fidkowski_10} and multi-layer systems~\cite{yoshida2017}. In the latter, the $\mathbb{Z} \times \mathbb{Z}$ classification is reduced to a $\mathbb{Z} \times \mathbb{Z}_8$ classification, indicating that the topological phases with a mirror Chern number of $8 \mathbb{Z}$ are fragile against interactions. This reduction of topology could be observed in the odd-parity $d$-wave superconducting phase for quad-layer systems, as the mirror Chern number is predicted to be $\nu_{\rm M}=8$~\cite{yoshida2017}.

Even when no in-plane mirror symmetry exists, such as is the case in the nonsymmorphic CeRh$_2$As$_2$ crystal structure, a topological crystalline phase can be defined: For the glide-plane invariant subspaces, in other words for $k_z = 0$ or $\pi$, we can define a 1D invariant for the special direction $k_x + k_y = 0$, which is a strong topological index based on the K-theory classification~\cite{shiozaki2016}. For CeRh$_2$As$_2$, this could result in a topologically non-trivial phase with zero-energy bound states for appropriate crystal terminations~\cite{nogaki:2021}.

\subsubsection{Spontaneous TRS breaking and Weyl superconductivity}
The staggered-Rashba systems with four-fold rotation axis are not expected to break time-reversal symmetry spontaneously similar to the globally mirror-symmetry-broken systems~\cite{scheurer:2017}. However, no such restrictions exist for systems with six-fold symmetry and, in particular, the $C_2$-lacking hexagonal superconductors. 
For both SrPtAs and 4Hb-TaS$_2$, time-reversal-symmetry breaking was indeed found upon entering the superconducting phase in $\mu$SR experiments~\cite{biswas:2013, ribak:2020}. These measurements suggest that both superconductors could exhibit chiral phases of $d+id$ symmetry. While no 3D topological invariant exists since TRS is broken, a Chern number can be defined for each $k_z$ slice. If this Chern number changes as a function of $k_z$, the system can realize a Weyl superconductor with Majorana Fermi arcs on the surfaces as was argued in the case of SrPtAs~\cite{fischer:2014a,fischer:2015a}.

\section{Outlook}\label{sec:finish}

A key ingredient to observing the novel physics of locally noncentrosymmetric superconductors is ensuring that the ASOC is larger than the intersublattice coupling. Earlier works suggested that this could be implemented by constructing appropriate heterostructures. More recently, a similar strategy has been adopted to create clean crystalline materials. In one example, Ba$_6$Nb$_{11}$S$_{28}$ \cite{devarakonda_2020}, spacer layers have successfully been added between layers of NbS$_2$ to form an ideal quasi-two-dimensional material in which ASOC is enhanced. Another approach is to generate ASOC through internal degrees of freedom with opposite inversion symmetry. An example in this case is monolayer WTe$_2$, in which a large ASOC is generated by coupling Te p and W d-orbitals \cite{Xie_2020} which accounts for the large observed in-plane upper critical field. 

Another strategy to enhance the ASOC is to exploit nonsymmorphic space groups. Indeed, such space groups even allow for new aspects of locally noncentrosymmetric superconductors to be revealed. As illustrated in {\bf Figure~\ref{fig:examples}b}, many locally inversion-symmetry-broken crystals have a nonsymmorphic space group. A characteristic property of nonsymmorphic systems are band degeneracies on the glide invariant planes or screw invariant lines, in addition to the usual spin degeneracy~\cite{Bradley}. This degeneracy may enforce a vanishing intersublattice coupling in the normal-state Hamiltonian. As discussed above, a vanishing (or small) intersublattice coupling enhances the role of ASOC, and hence the effects of local inversion-symmetry breaking. For superconductors, this results in increased residual spin susceptibility at zero temperature and paramagnetically-limiting upper critical fields. The field-induced odd-parity superconducting state discussed in Sec.~\ref{sec:parity-transition} is further stabilized~\cite{Sumita_2016,Cavanagh_2022}, which is relevant to the physics of CeRh$_2$As$_2$~\cite{Cavanagh_2022}.
The vanishing intersublattice coupling further leads to a striking property of locally noncentrosymmetric superconductors: line nodes in the gap of odd-parity superconductors are allowed~\cite{Micklitz-Norman2009}, providing a counter example to Blount's theorem~\cite{Blount1985}. As demonstrated for a model of UPt$_3$, the appearance of line nodes can be intuitively understood by decomposing the system into a pair of noncentrosymmetric systems using the vanishing intersublattice coupling~\cite{Yanase_UPt3_2016}. 

Apart from a vanishing intersublattice coupling, the nonsymmorphic space groups result in unusual properties in the Bloch wave functions. In particular, the space-group symmetries cause a M\"obius structure in the Brillouin zone, in the sense that the Bloch wave functions show a $4\pi$ periodicity, instead of the usual $2\pi$ periodicity. As a consequence, the topological phases of insulators/semiconductors/superconductors are enriched~\cite{shiozaki2016}, as discussed for CeRh$_2$As$_2$ in Sec.~\ref{sec:TCSC} as well as for UPt$_3$~\cite{yanaseUPt32017} and UCoGe~\cite{daido2019}. Such topological superconducting phases are unique in lattice systems in the sense that there is no counter part in the continuum system.

One final promising direction is the novel role of spin-fluctuations in locally noncentrosymmetric crystals in stabilizing unconventional superconducting states. One material that illustrates this is nonsymmorphic Sr$_2$IrO$_4$, where intersublattice spin-singlet superconductivity is suppressed by the ASOC~\cite{fischer:2011b}. As a consequence, spin-triplet superconductivity can be stabilized~\cite{ishizuka2018}. In contrast to the ordinary spin-triplet pairing mechanism due to ferromagnetic fluctuations, the ASOC due to the local inversion-symmetry breaking and the nonsymmorphic space group conspire such that {\it antiferromagnetic} spin fluctuations stabilize a spin-triplet state with non-trivial topology~\cite{ishizuka2018}. Strongly correlated electrons in locally noncentrosymmetric crystals, thus, provide a route to realizing spin-triplet topological superconductivity without ferromagnetic fluctuations.

\section*{DISCLOSURE STATEMENT}
The authors are not aware of any affiliations, memberships, funding, or financial holdings that
might be perceived as affecting the objectivity of this review. 

\section*{ACKNOWLEDGMENTS}
We thank Manuel Brando, Philip Brydon, David Cavanagh, Elena Hassinger, Seunghyun Khim, Andy Mackenzie, Daisuke Maruyama, Igor Mazin, David Möckli, Titus Neupert, Kosuke Nogaki, Aline Ramires, Erik Schertenleib, Tatsuya Shishidou, Anastasiia Skurativska, Michael Weinert, Tomohiro Yoshida, and Tsuneya Yoshida. D.F.A. was supported by the US Department of Energy, Office of Basic Energy Sciences, Division of Materials Sciences and Engineering, under award DE-SC0021971. M.S. is grateful for financial support from Swiss National Science Foundation (SNSF) through Division II (No. 184739). 
Y.Y. was supported by JSPS KAKENHI (Grants No. JP18H05227, No. JP18H01178, No. JP20H05159) and SPIRITS 2020 of Kyoto University.

\bibliographystyle{ar-style4.bst}

\bibliography{refs}

\end{document}